\documentclass[aps,prd,superscriptaddress,amsmath,twocolumn,10pt]{revtex4}

\usepackage{color,hangcaption,hhline,psfrag,rotating,amssymb}
\usepackage[hang,nooneline]{subfigure}
\usepackage{dcolumn}
\usepackage{verbatim}

\newcommand{\be}{\begin{equation}}
\newcommand{\ee}{\end{equation}}
\newcommand{\bea}{\begin{eqnarray}}
\newcommand{\eea}{\end{eqnarray}}

\begin{document}
\title{B-meson decay constants from improved lattice NRQCD and physical u, d, s and c sea quarks}

\author{R.~J.~Dowdall}
\email[]{R.J.Dowdall@damtp.cam.ac.uk}
\affiliation{DAMTP, University of Cambridge, Wilberforce Road, Cambridge CB3 0WA, UK}
\affiliation{SUPA, School of Physics and Astronomy, University of Glasgow, Glasgow, G12 8QQ, UK}
\author{C.~T.~H.~Davies}
\email[]{Christine.Davies@glasgow.ac.uk}
\affiliation{SUPA, School of Physics and Astronomy, University of Glasgow, Glasgow, G12 8QQ, UK}
\author{R.~R.~Horgan}
\affiliation{DAMTP, University of Cambridge, Wilberforce Road, Cambridge CB3 0WA, UK}
\author{C.~J.~Monahan}
\affiliation{Physics Department, College of William and Mary, Williamsburg, Virginia 23187, USA}
\author{J.~Shigemitsu}
\affiliation{Physics Department, The Ohio State University, Columbus, Ohio 43210, USA}

\collaboration{HPQCD collaboration}
\homepage{http://www.physics.gla.ac.uk/HPQCD}

\date{\today}

\begin{abstract}
We present the first lattice QCD calculation of the decay constants $f_B$ and $f_{B_s}$ with physical light quark masses. We use configurations generated by the MILC collaboration including the effect of $u$, $d$, $s$ and $c$ HISQ sea quarks 
 at three lattice spacings and with three $u/d$ quark mass values going down to the physical value. We use improved NRQCD for the valence $b$ quarks. 
Our results are $f_B = 0.186(4)$ GeV, $f_{B_s} = 0.224(5)$ GeV, $f_{B_s}/f_B = 1.205(7)$ and $M_{B_s}-M_B=85(2)$ MeV, superseding earlier results with NRQCD $b$ quarks. 
We discuss the implications of our results for the Standard Model 
rates for $B_{(s)} \rightarrow \mu^+\mu^-$ 
and $B \rightarrow \tau \nu$. 
\end{abstract}


\maketitle

%
\section{Introduction}
\label{sec:intro}
%
The $B$ and $B_s$ decay constants are key hadronic parameters in 
the Standard Model (SM) rate for $B_{(s)} \rightarrow \mu^+\mu^-$ 
and $B/B_s$ oscillations, with 
the $B$ meson decay constant also determining the 
rate for $B \rightarrow \tau \nu$. The combination of experiment 
and theory for these processes provides important constraints on 
CKM unitarity~\cite{Laiho:2012ss} and the search 
for new physics, but the strength of the constraints is typically limited 
by the errors on the hadronic parameters.  

The decay constants can only be 
determined accurately from lattice QCD calculations. 
Several methods 
have been developed for this~\cite{Davies:2012qf}, with errors decreasing 
over the years as calculations have improved. 
Here we provide a step 
change in this process, giving the first results for $f_B$ and $f_{B_s}$ 
that include physical $u/d$ quark masses, obviating the 
need for a chiral extrapolation. As a result of this and other improvements 
described below, we have signficantly improved accuracy on $f_{B_s}/f_B$ 
over previous calculations. The implications of our result are discussed 
in the Conclusions.

%
\section{Lattice calculation}
\label{sec:lattice}
%

\begin{table}
\caption{
Details of the gauge ensembles used in this calculation. $\beta$ is the gauge coupling, $a_{\Upsilon}$ is the lattice spacing as 
determined by the $\Upsilon(2S-1S)$ splitting in \cite{Dowdall:2011wh}, where the three errors are statistics, NRQCD systematics and experiment. 
$am_l,am_s$ and $am_c$ are the sea quark masses, $L \times T$ gives the spatial and temporal extent of the lattices and $n_{{\rm cfg}}$ is the number of configurations in each ensemble. 
The ensembles 1,2 and 3 will be referred to as ``very coarse'', 4,5 and 6 as ``coarse'' and 7,8 as ``fine''. 
}
\label{tab:gaugeparams}
\begin{ruledtabular}
\begin{tabular}{lllllllll}
Set & $\beta$ & $a_{\Upsilon}$ (fm) 	& $am_{l}$ & $am_{s}$ & $am_c$ & $L \times T$ & $n_{{\rm cfg}}$  \\
\hline
1 & 5.8 & 0.1474(5)(14)(2)  & 0.013   & 0.065  & 0.838 & 16$\times$48 & 1020 \\
2 & 5.8 & 0.1463(3)(14)(2)  & 0.0064  & 0.064  & 0.828 & 24$\times$48 & 1000 \\
3 & 5.8 & 0.1450(3)(14)(2)  & 0.00235 & 0.0647 & 0.831 & 32$\times$48 & 1000 \\
\hline
4 & 6.0 & 0.1219(2)(9)(2)   & 0.0102  & 0.0509 & 0.635 & 24$\times$64 & 1052 \\
5 & 6.0 & 0.1195(3)(9)(2)   & 0.00507 & 0.0507 & 0.628 & 32$\times$64 & 1000 \\
6 & 6.0 & 0.1189(2)(9)(2)   & 0.00184 & 0.0507 & 0.628 & 48$\times$64 & 1000 \\
\hline
7 & 6.3 & 0.0884(3)(5)(1)   & 0.0074  & 0.037  & 0.440 & 32$\times$96 & 1008 \\
8 & 6.3 & 0.0873(2)(5)(1)   & 0.0012  & 0.0363 & 0.432 & 64$\times$96 & 621 \\
\end{tabular}
\end{ruledtabular}
\end{table}

\begin{table}
\caption{ 
\label{tab:params}
Parameters used for the valence quarks. $am_b$ is the bare $b$ quark mass in lattice units, $u_{0L}$ is the Landau link value used for tadpole-improvement,
and $am_l^{\rm val}$, $am_s^{\rm val}$ are the HISQ light and strange quark masses.
}
\begin{ruledtabular}
\begin{tabular}{lllll}
Set & $am_b$ & $u_{0L}$ & $am_l^{\rm val}$ & $am_s^{\rm val}$  \\
\hline
1 & 3.297 & 0.8195   & 0.013   & 0.0641 \\
2 & 3.263 & 0.82015  & 0.0064  & 0.0636 \\
3 & 3.25  & 0.819467 & 0.00235 & 0.0628 \\
\hline
4 & 2.66 & 0.834    & 0.01044 & 0.0522  \\ 
5 & 2.62 & 0.8349   & 0.00507 & 0.0505  \\ 
6 & 2.62 & 0.834083 & 0.00184 & 0.0507  \\
\hline
7 & 1.91 & 0.8525   & 0.0074  & 0.0364  \\ 
8 & 1.89 & 0.851805 & 0.0012  & 0.0360  
\end{tabular}
\end{ruledtabular}
\end{table}

We use eight ensembles of `second-generation' gluon field configurations recently generated by the MILC collaboration \cite{Bazavov:2010ru,MILC:2012uw}, with $N_f=2+1+1$ Highly Improved Staggered Quarks (HISQ) \cite{Follana:2006rc} in the sea. To control discretisation effects, we use three lattice spacings ranging from 0.15 fm to 0.09 fm and light to strange mass ratios of $m_l/m_s\sim0.2,0.1,0.037$.
Details of the ensembles are shown in table \ref{tab:gaugeparams}.
The lattice spacings of five of the ensembles were determined using the $\Upsilon(2S-1S)$ splitting in~\cite{Dowdall:2011wh} where details, including a discussion of the systematic errors, can be found. 
The lattice spacing values of the additional ensembles (sets 3, 6 and 8) are determined in the same way.
The valence part of the calculation uses lattice NonRelativistic QCD (NRQCD) \cite{Lepage:1992tx,Thacker:1990bm,Gray:2005ur} for the $b$ quarks; the action is described in detail in  \cite{Dowdall:2011wh}. 
It includes a number of improvements over earlier calculations, 
in particular one-loop radiative corrections (beyond tadpole-improvement) 
to most of 
the coefficients of the $\mathcal{O}(v_b^4)$ relativistic correction terms. 
This action has been shown to give excellent agreement with experiment in recent calculations of the bottomonium \cite{Dowdall:2011wh,Daldrop:2011aa} and $B$-meson spectrum \cite{Dowdall:2012ab}. 
We are now building on previous calculations with the tree level 
NRQCD action~\cite{Gamiz:2009ku, Gregory:2010gm, Na:2012kp} to extend this to $B$-meson decay constants.
The $b$ quark mass is tuned, giving the values in Table~\ref{tab:params},  by
fixing the spin-averaged kinetic mass with the $\Upsilon/\eta_b$ masses. 

The HISQ valence light quark masses are taken to be equal to the sea mass except on set 4 where there is a slight discrepancy. 
The $s$ quark is tuned using the $\eta_s$ meson ($M_{\eta_s} =$ 0.6893(12) GeV~\cite{Dowdall:2011wh}). 
Values very close to the sea $s$ masses 
are found, meaning that partial quenching effects will be small.

To improve the statistical precision of the correlators, 
we take $U(1)$ random noise sources for the valence quarks 
using the methods developed in~\cite{Gregory:2010gm}.
Along with the point source required for the matrix element, 
we include gaussian smearing functions for the $b$ quark 
source with two different widths. We include 16 time sources with 
$b$ quarks propagating both forward and backward in time on 
each configuration. 
We checked the statistical independence of results
using a blocked autocorrelation function~\cite{Dowdall:2011wh}. 
Even on the finer physical 
point ensembles, the correlations are very small 
between adjacent configurations and the integrated autocorrelation time is consistent with one.

The decay constant is defined from 
$\langle 0 | A_0  |B_q \rangle_{\rm QCD} = M_{B_q} f_{B_q}$, but 
the quantity that we extract directly from the amplitude of 
our correlator fits is $\Phi_{B_q}=\sqrt{M_{B_q}} f_{B_q}$; we convert 
to $f_{B_q}$ at the end. 
For NRQCD, 
the full QCD matrix element is constructed from 
effective theory currents arranged in powers of $1/m_b$.
For $A_0$ we consider the following currents, 
made from heavy quark $\Psi_Q$ and light quark fields $\Psi_q$:  
\begin{eqnarray}
J_0^{(0)} &=& \bar{\Psi}_q \gamma_5 \gamma_0 \Psi_Q \\
J_0^{(1)} &=& \frac{-1}{2m_b} \bar{\Psi}_q \gamma_5 \gamma_0 \gamma \cdot \nabla \Psi_Q \\
J_0^{(2)} &=& \frac{-1}{2m_b} \bar{\Psi}_q \gamma \cdot \overleftarrow{\nabla}   \gamma_5 \gamma_0 \Psi_Q . 
\end{eqnarray}
These currents are related to the full QCD current through
$\mathcal{O}\left(  \alpha_s, \alpha_s \Lambda_{\rm QCD}/m_b  \right)$ by
\begin{multline}
  \langle A_0 \rangle =
 (1+ \alpha_s z_0) \left(  \langle J_0^{(0)} \rangle  \right.\\
 +  (1+ \alpha_s z_1  ) \langle J_0^{(1)} \rangle 
  + \left.  \alpha_s  z_2   \langle J_0^{(2)} \rangle
\right)
\label{eq:renorm}
\end{multline}

One-loop coefficients were calculated in~\cite{Monahan:2012dq}. Here 
we re-order the perturbation series to make the process of renormalisation 
clearer. The $z_i$ depend on $am_b$ and are given 
in Table~\ref{tab:zresults} 
for the range of masses needed here.
We see that the one-loop renormalisation of the 
tree-level current, $J_0^{(0)}+J_0^{(1)}$, is tiny~\footnote{This agrees 
with expectations from~\cite{Harada:2001fi, Bazavov:2011aa} in which the heavy-light renormalisation constant is perturbatively very close to
the product of the square roots of the 
renormalisation of the local temporal vector current 
for heavy-heavy and light-light. 
Here the corresponding heavy-heavy current is conserved~\cite{Boyle:2000fi} and 
the light-light current has a very small renormalisation~\cite{Donald:2012ga}. This
will be discussed further elsewhere. }. 
$z_0$ includes the effect of mixing between $J_0^{(0)}$ 
and $J_1^{(1)}$ at one-loop.
We evaluate the renormalisation of Eq.~\ref{eq:renorm} using $\alpha_s$ 
in the V-scheme at scale $q=2/a$. Values for $\alpha_s$ are obtained by 
running down from $\alpha_s^{\overline{\rm MS}}(M_Z)=0.1184$~\cite{McNeile:2010ji} 
and range from 0.285 to 0.314.

\begin{table}
\caption{\label{tab:zresults} 
Coefficients for the perturbative matching of the 
axial vector current (Eq.~\ref{eq:renorm}). 
$z_0 = \rho_0-\zeta_{10}$, $z_1 = \rho_1 - z_0$, 
$z_2 = \rho_2$ from~\cite{Monahan:2012dq}.
}
\begin{ruledtabular}
\begin{tabular}{llllllll}
Set & $z_0$ & $z_1$ & $z_2$  \\
\hline \hline
1&0.024(2)  & 0.024(3)  & -1.108(4) \\
2&0.022(2)  & 0.024(3)  & -1.083(4) \\
3&0.022(1)  & 0.024(2)  & -1.074(4) \\
4&0.006(2)  & 0.007(3)  & -0.698(4) \\
5&0.001(2)  & 0.007(3)  & -0.690(4) \\
6&0.001(2)  & 0.007(2)  & -0.690(4) \\
7&-0.007(2) & -0.031(4) & -0.325(4) \\
8&-0.007(2) & -0.031(4) & -0.318(4) \\
\end{tabular}
\end{ruledtabular}
\end{table}

%
\section{Results}
\label{subsec:Results}
%
\begin{table}
\begin{ruledtabular}
\caption{\label{tab:ampresults} 
Raw lattice amplitudes for $B_s$ and $B$ from each ensemble, errors are from statistics/fitting only. $a^{3/2}\Phi^{(0)}_q$  and $a^{3/2}\Phi^{(1)}_q$ are the leading amplitude and $1/m_b$ correction. 
}
\begin{tabular}{lllll}
Set &  $a^{3/2}\Phi^{(0)}_s$  & $a^{3/2}\Phi^{(1)}_s$  & $a^{3/2}\Phi^{(0)}$  & $a^{3/2}\Phi^{(1)}$     \\%
\hline \hline
1 & 0.3720(10) & -0.0300(3) & 0.3220(19) & -0.0260(3)\\ 
2 & 0.3644(6) & -0.0291(3) & 0.3093(11) & -0.0257(8)\\ 
3 & 0.3621(16) & -0.0288(2) & 0.2986(17) & -0.0237(4)\\ 
4 & 0.2733(4) & -0.0234(2) & 0.2373(9) & -0.0197(4)\\ 
5 & 0.2679(3) & -0.0234(1) & 0.2272(7) & -0.0197(3)\\ 
6 & 0.2653(2) & -0.0229(1) & 0.2193(8) & -0.0194(3)\\ 
7 & 0.1747(3) & -0.0170(1) & 0.1525(8) & -0.0146(6)\\ 
8 & 0.1694(3) & -0.0167(0) & 0.1386(5) & -0.0136(1)\\ 
\end{tabular}
\end{ruledtabular}
\end{table}

\begin{table}
\caption{\label{tab:enresults} 
Raw lattice energies from each ensemble, errors are from statistics/fitting only. $aM_{\pi}$ are the pion masses used in the chiral fits, $aE(B_s)$ and $aE(B)$ are the energies of the $B_s$ and $B$ meson. Results on sets 3, 6 and 8 are 
new, others are given in~\cite{Dowdall:2012ab}.
}
\begin{ruledtabular}
\begin{tabular}{llllllll}
Set & $aM_{\pi}$ & $aE(B_s)$ & $aE(B)$      \\
\hline \hline
3 & 0.10171(4) & 0.6067(7) & 0.5439(12)\\ 
6 & 0.08154(2) & 0.5158(1) & 0.4649(6)\\ 
8 & 0.05718(1) & 0.4025(2) & 0.3638(5)\\ 
\end{tabular}
\end{ruledtabular}
\end{table}

We fit heavy-light meson correlators with both $J_0^{(0)}$ and 
$J_0^{(1)}$ operators at the sink simultaneously using a 
multi-exponential Bayesian fitting procedure~\cite{gplbayes}. 
The $B$ and $B_s$ are fit separately; 
priors used in the fit are described in~\cite{Dowdall:2012ab}. 
The amplitudes and energies from the fits are given in Tables \ref{tab:ampresults} and \ref{tab:enresults}. 
$a^{3/2}\Phi^{(0)}_q$ is the matrix element of the leading current $J_0^{(0)}$ 
and $a^{3/2}\Phi^{(1)}_q$ that of $J_0^{(1)}$ and $J_0^{(2)}$, whose matrix elements are equal at zero meson momentum.
Notice that the statistical errors in $\Phi$ do not increase on the physical point 
lattices, because they have such large volumes. 

We take two approaches to the analysis. 
The first is to perform a simultaneous chiral 
fit to all our results for $\Phi,\Phi_s,\Phi_s/\Phi$ and $M_{B_s}-M_B$ using 
$SU(2)$ chiral perturbation theory. 
The second is to study only the physical 
$u/d$ mass results as a function of lattice spacing. 

For the chiral analysis we use the same formula and priors for 
$M_{B_s}-M_B$ as in~\cite{Dowdall:2012ab}. 
Pion masses used in the fits are listed in Table \ref{tab:enresults} and the chiral logarithms, $l(M^2_\pi)$, include 
the finite volume corrections computed in~\cite{Bernard:2001yj} which 
have negligible effect on the fit.
For the decay constants the chiral formulas, 
including analytic terms up to $M_\pi^2$ and the 
leading logarithmic behaviour, are (see e.g. \cite{Albertus:2010nm}):
\begin{eqnarray}
\Phi_s &=& \Phi_{s0}(1.0 + b_s M_\pi^2/ \Lambda_\chi^2 )\\
\Phi &=& \Phi_{0} \left( 1.0 + b_l \frac{M_\pi^2}{\Lambda_\chi^2} + \frac{1+3g^2}{2 \Lambda_\chi^2}  \left( -\frac{3}{2}l(M_\pi^2) \right)  \right)
\end{eqnarray}
The coefficients of the analytic terms $b_s,b_l$ are given 
priors 0.0(1.0) and $\Phi_0, \Phi_{s0}$ 
have 0.5(5).
To allow for discretisation errors each fit formula is multiplied by 
$
    (1.0 + d_1(\Lambda a)^2 + d_2(\Lambda a)^4 )
$,
with $\Lambda=0.4$ GeV. 
We expect discretisation effects to be very similar for $\Phi$ and $\Phi_s$ 
and so we take the $d_i$ to be the same, but differing from the $d_i$ 
used in the $M_{B_s}-M_B$ fit.
Since all actions used here are accurate through $a^2$ at tree-level, the prior on $d_1$ is taken to be 0.0(3) whereas $d_2$ is 0.0(1.0).
The $d_i$ are allowed to have mild $m_b$ dependence as in \cite{Dowdall:2012ab}.
The ratio $\Phi_s/\Phi$ is allowed additional light quark 
mass dependent discretisation errors that could arise, for example, 
from staggered taste-splittings.

\begin{table}[ht]
\begin{ruledtabular}
\begin{tabular}{lcccc}
Error \%	& $\Phi_{B_s}/\Phi_B$ & $M_{B_s}-M_B$ & $\Phi_{B_s}$ & $\Phi_{B}$ \\
\hline
EM: 		& 0.0 	& 1.2 	& 0.0 & 0.0 \\ 
$a$ dependence:	& 0.01 	& 0.9 	& 0.7 & 0.7 \\ 
chiral: 	& 0.01 	& 0.2 	& 0.05 & 0.05\\ 
g: 		& 0.01 	& 0.1 	& 0.0 & 0.0 \\ 
stat/scale: 	& 0.30  & 1.2 	& 1.1 & 1.1 \\ 
operator: 	& 0.0   & 0.0 	& 1.4 & 1.4 \\ 
relativistic: 	& 0.5   & 0.5 	& 1.0 & 1.0 \\ 
\hline
total: 		& 0.6  & 2.0   & 2.0  & 2.1 \\ 
\hline
\end{tabular}
\end{ruledtabular}
\caption{\label{tab:errors} 
Full error budget from the chiral fit as a percentage of the final answer. }
\end{table}

\begin{figure}
\includegraphics[width=\hsize]{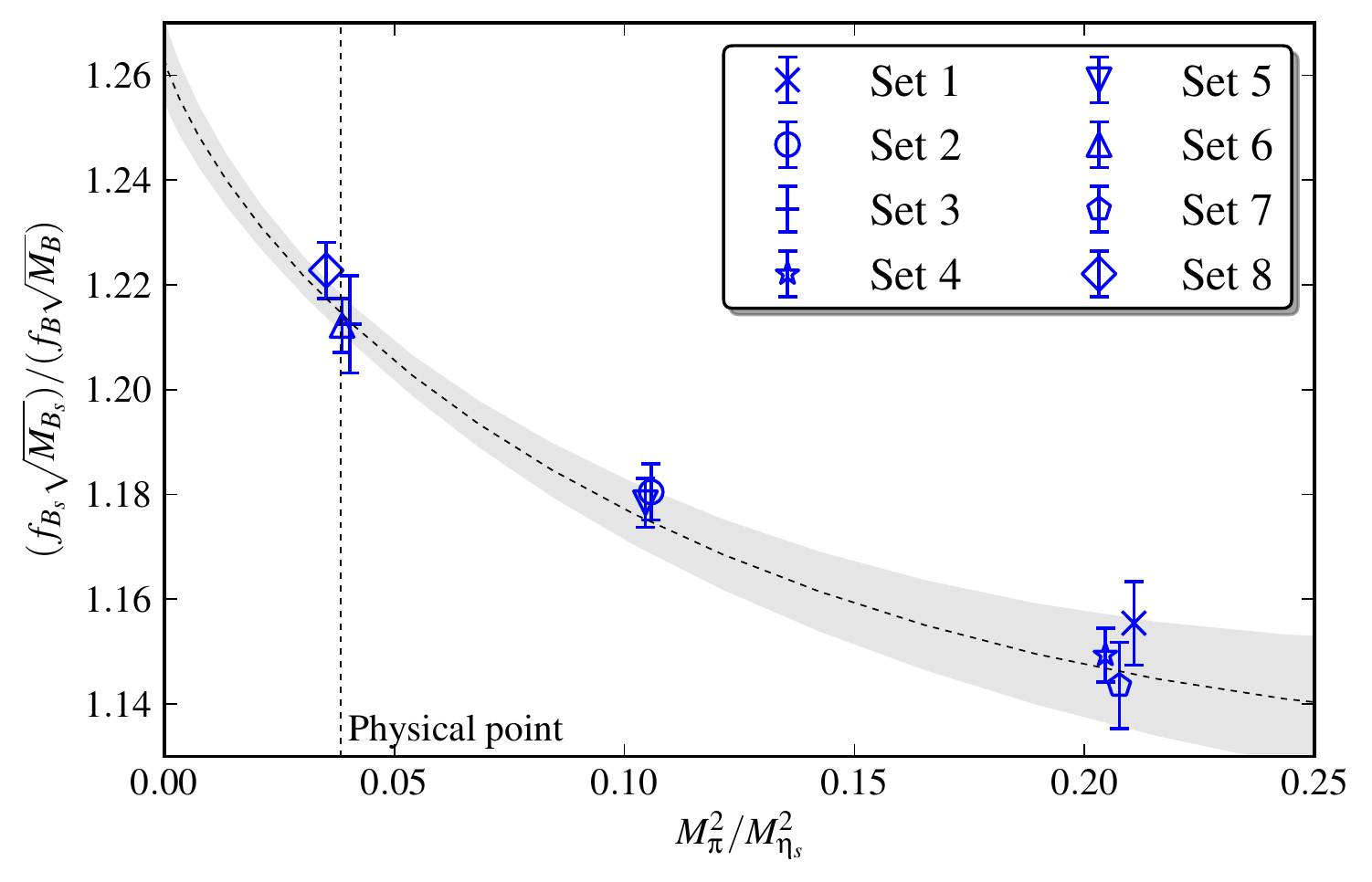}
\caption{\label{fig:fBsfB}Fit to the decay constant ratio $\Phi_{B_s}/\Phi_B$. The fit result is shown in grey and errors include statistics, and chiral/continuum fitting.}
\end{figure}

\begin{figure}
\includegraphics[width=\hsize]{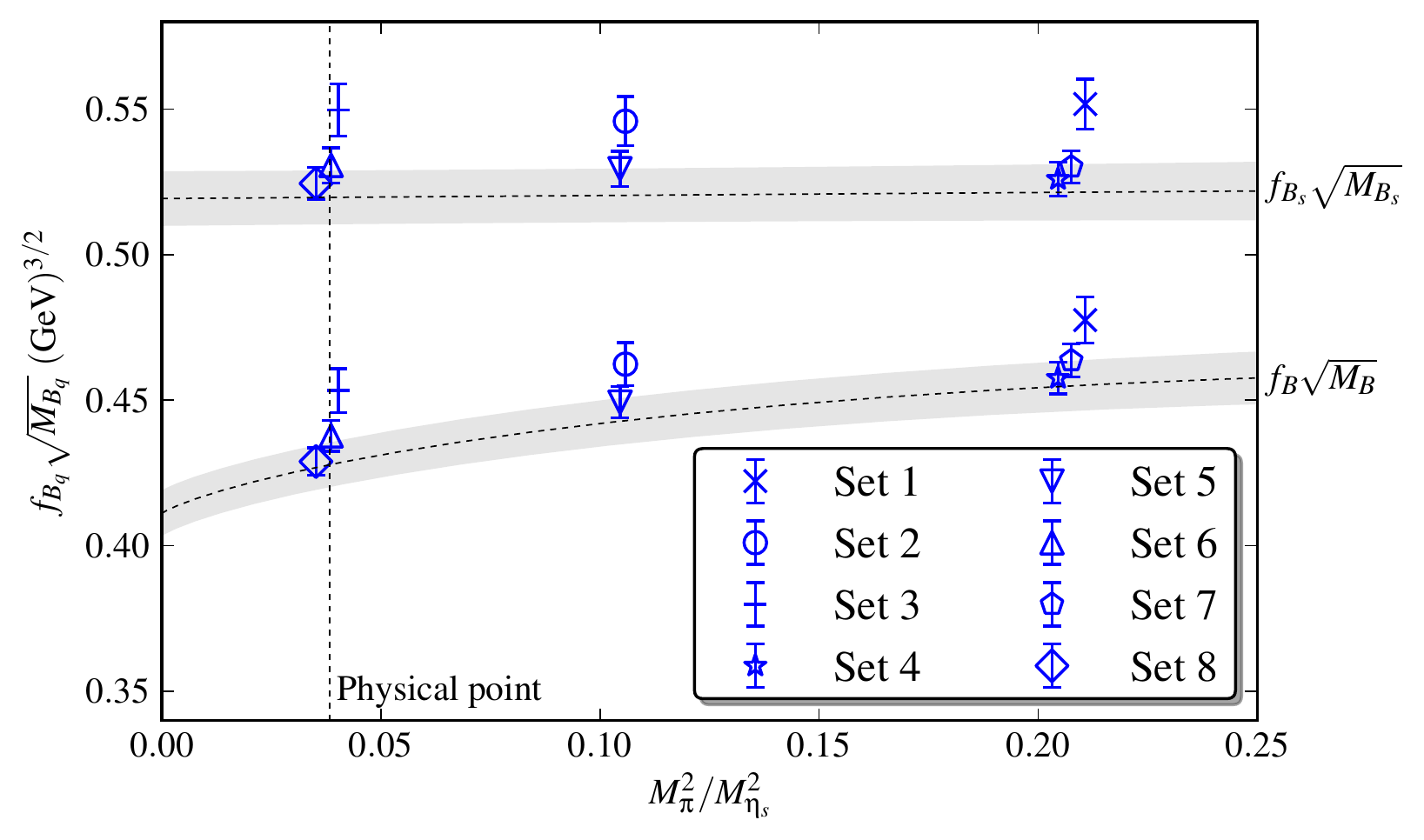}
\caption{\label{fig:fBs}Fit to the decay constants $\Phi_{B_s}$ and $\Phi_{B}$. 
Errors on the data points include statistics/scale only. 
The fit error, in grey, includes chiral/continuum fitting and perturbative errors.}
\end{figure}

The results of the decay constant chiral fits are plotted in 
Figs.~\ref{fig:fBsfB} and~\ref{fig:fBs}. 
Extrapolating to the physical point appropriate to $m_l=(m_u+m_d)/2$ 
in the absence of electromagnetism, i.e. 
$M_{\pi}=M_{\pi^0}$,
 we find $\Phi_{B_s}= 0.520(11) \ {\rm GeV^{3/2}}$, $\Phi_B= 0.428(9) \ {\rm GeV^{3/2}}$,
$\Phi_{B_s} / \Phi_B=  1.215(7)$.
For $M_{B_s}-M_B$ we obtain 86(1) MeV, 
in agreement with the result of~\cite{Dowdall:2012ab}. 

\begin{figure}
\includegraphics[width=\hsize]{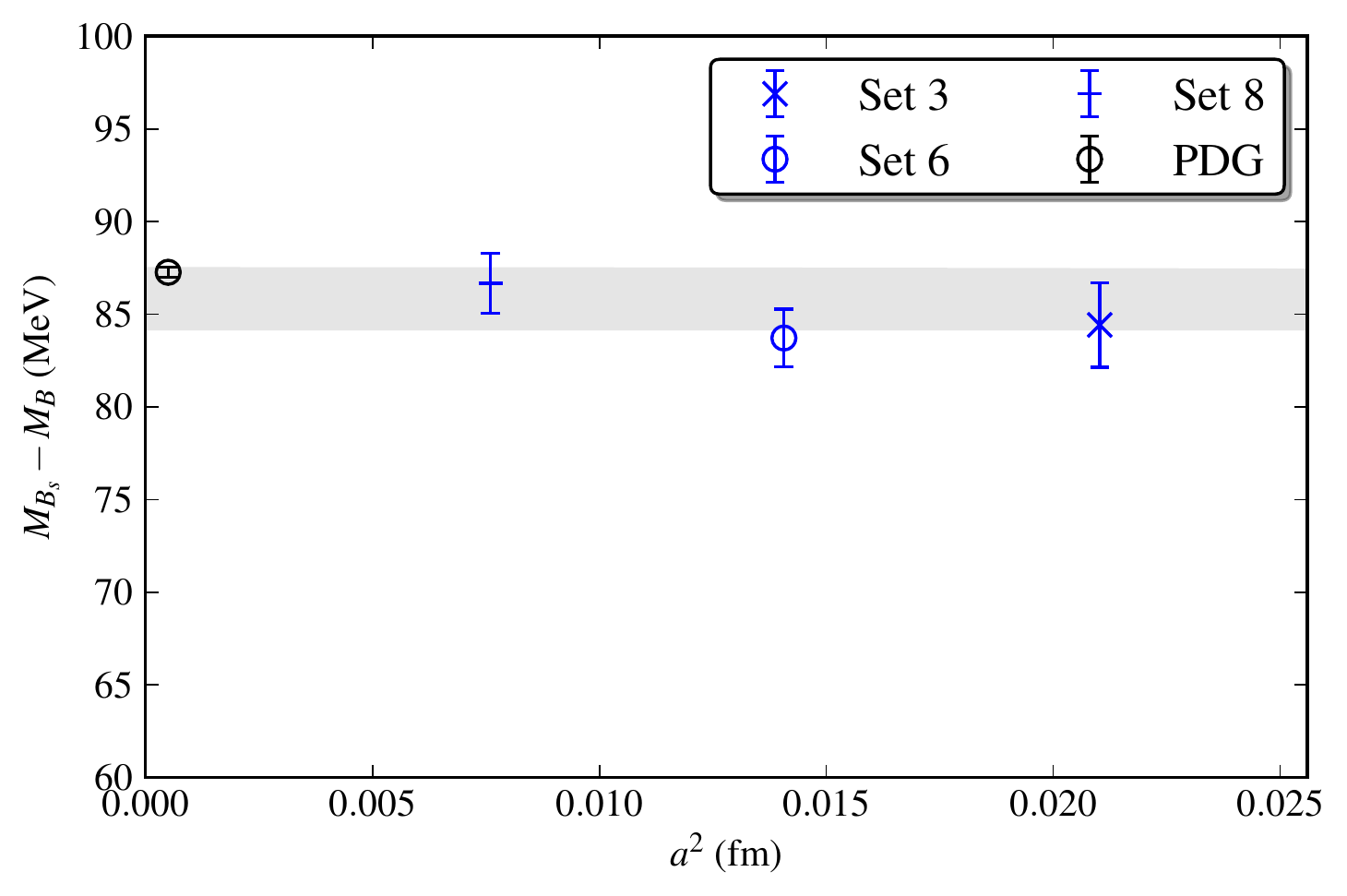}
\caption{\label{fig:MBsMB}Fit to the mass difference $M_{B_s}-M_B$ on the three physical point ensembles only. Errors on data points include statistics and scale, the fit error is shown in grey. An electromagnetic correction of -1(1) MeV has been applied to the lattice results and the fit to allow comparison with experiment.}
\end{figure}

\begin{figure}
\includegraphics[width=\hsize]{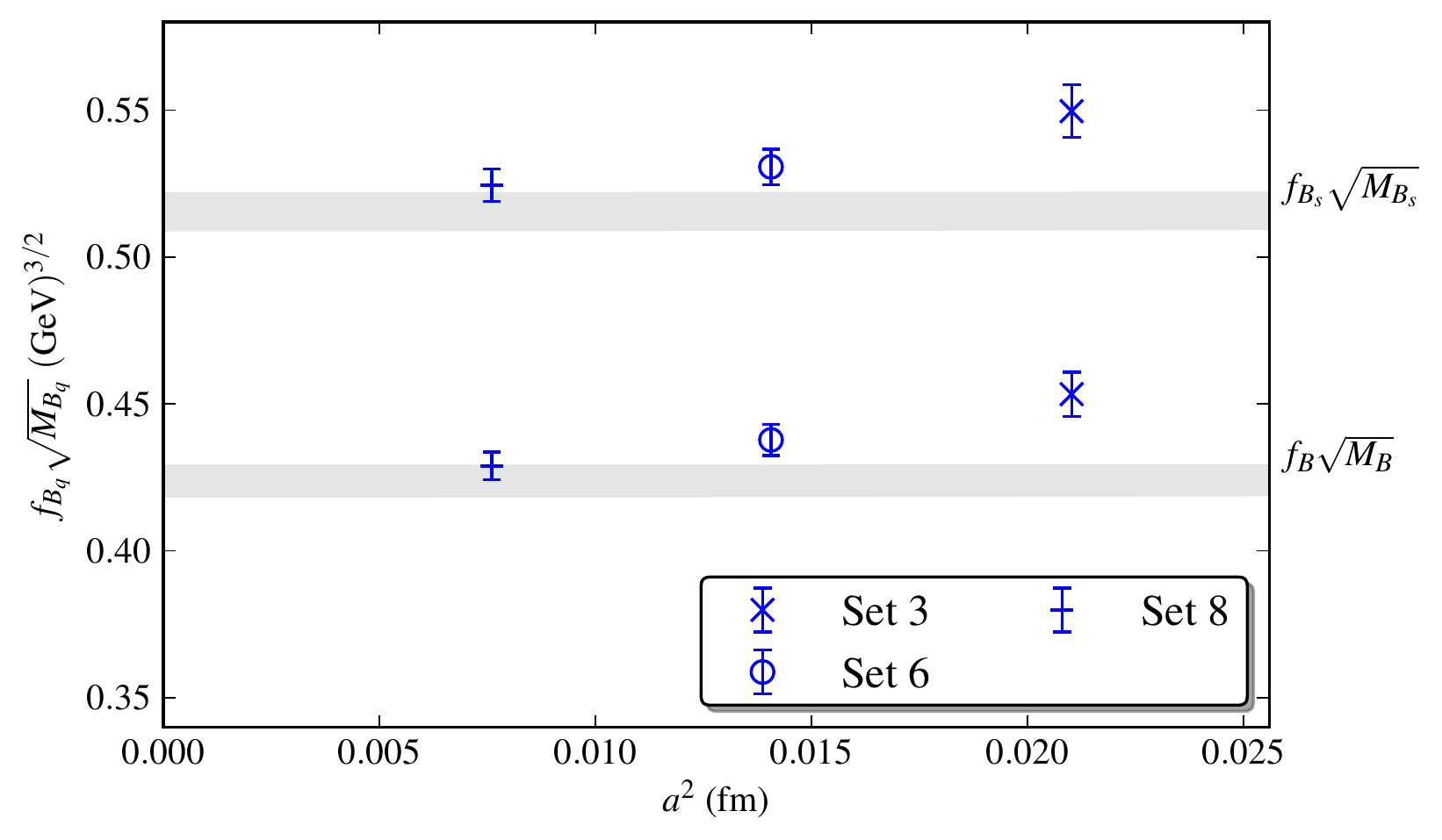}
\caption{\label{fig:fBs-physpt}Fit to the decay constants $\Phi_{B_s}$ and $\Phi_{B}$ on the three physical point ensembles only. 
Errors on the data points include statistics/scale only. 
The fit error includes chiral/continuum fitting and perturbative errors.}
\end{figure}

Figs~\ref{fig:MBsMB} and~\ref{fig:fBs-physpt} 
show the results of fitting $M_{B_s}-M_B$ and decay constants from 
the physical point ensembles only, and allowing only the mass dependent discretisation terms above. 
The results are $\Phi_{B_s}=  0.515(8) \ {\rm GeV^{3/2}}$, $\Phi_B= 0.424(7) \ {\rm GeV^{3/2}}$,
$\Phi_{B_s} / \Phi_B=  1.216(7)$ and $M_{B_s}-M_B = 87(1)$ MeV. 
Results and errors agree well between the two methods and we take the central values from the chiral 
fit as this allows us to interpolate to the correct pion mass. 

Our error budget is given in Table~\ref{tab:errors}. 
The errors that are estimated directly from 
the chiral/continuum fit are those from statistics, the lattice spacing 
and $g$ and other chiral fit parameters. 
The two remaining sources of error in the decay constant are 
missing higher order corrections in 
the operator matching and relativistic corrections to the 
current. We estimate the operator matching error by allowing 
in our fits for an $am_b$-dependent $\alpha_s^2$ correction 
to the renormalisation in Eq.~\ref{eq:renorm} with 
prior on the coefficient of 0.0(2) i.e. ten times the size 
of the one-loop correction, $z_0$. This error cancels 
in the ratio $f_{B_s}/f_B$. We also allow for $\alpha_s^2$ 
corrections multiplying $J_0^{(1,2)}$ with coefficient 0.0(1.0). 
The matrix element of $J_0^{(1)}$ is about 10\% of $J_0^{(0)}$ from 
Table~\ref{tab:ampresults}. 
Missing current corrections at the next order in $1/m_b$  will 
be of size $(\Lambda_{\rm QCD}/m_b)^2\simeq 0.01$ which we 
take as an error.
Finally, we estimated in~\cite{Dowdall:2012ab} that to correct for missing electromagnetic effects, 
$M_{B_s}-M_B$ should be shifted by -1(1) MeV.

Using the PDG masses $M_{B_l}=(M_{B^0}+M_{B^\pm})/2 = 5.27942(12)$ GeV and $M_{B_s}=$ 5.36668(24) GeV~\cite{pdg} 
to convert $\Phi_q$ to $f_{B_q}$ our final results are:
\begin{eqnarray}
\label{eq:fresults}
f_B &=& 0.186(4) {\rm \ GeV} \\
f_{B_s} &=&  0.224(5) {\rm \ GeV} \nonumber \\
f_{B_s}/f_B &=& 1.205(7) \nonumber \\
M_{B_s} - M_B &=&  85(2)  {\rm \ MeV}.\nonumber 
\end{eqnarray}
For the $B$ meson decay constant we need to distinguish between 
$f_{B_d}$ and $f_{B_u}$. Since sea quark mass effects are much smaller 
than valence mass effects we simply do this by extrapolating $\Phi_{B_s}$ and $\Phi_B$ 
to values of $M_{\pi}^2$ corresponding to fictitious mesons made purely 
of $u$ or $d$ quarks using $m_u/m_d=0.48(10)$~\cite{pdg}. 
This gives: 
\begin{eqnarray}
\label{eq:fmoreres}
f_{B_s}/f_{B^+} = 1.217(8)&;& \, f_{B_s}/f_{B^0} = 1.194(7) \nonumber \\
f_{B^+} = 0.184(4) {\rm \ GeV}&;& \, f_{B^0} = 0.188(4) {\rm \ GeV} 
\end{eqnarray}

%
\section{Conclusions}
\label{sec:conclusions}
%

Our results agree with but improve substantially on two earlier 
results using nonrelativistic 
approaches for the $b$ quark and multiple lattice spacing 
values on $N_f=2+1$ ensembles using asqtad sea quarks.  
These were: $f_{B_s} = 228(10)$ MeV, $f_{B_s}/f_B = 1.188(18)$  
(NRQCD/HISQ)~\cite{Na:2012kp} and
$f_{B_s}=242.0(9.5)$ MeV and $f_{B_s}/f_{B^+}=1.229(26)$ (Fermilab/asqtad)~\cite{Bazavov:2011aa}. 
We also agree well (within the 2\% errors) 
with a previous result for $f_{B_s}$ of 225(4) MeV 
obtained using a relativistic 
(HISQ) approach to $b$ quarks on very fine $N_f=2+1$ 
lattices~\cite{McNeile:2011ng}. Our simultaneous determination of 
$M_{B_s}-M_{B}$ to 2\% agrees with experiment (87.4(3) MeV~\cite{pdg}).

We can determine new lattice `world-average' error-weighted 
values by combining 
our results in Eq.~\ref{eq:fresults} with the independent 
results of~\cite{Bazavov:2011aa} and~\cite{McNeile:2011ng}  
since effects from $c$ sea quarks, 
which they do not include, 
should be negligible~\cite{newfds}. The world averages 
are then: $f_{B_s} =$ 225(3) MeV 
and $f_{B_s}/f_{B^+} = 1.218(8)$ giving $f_{B^+} = $ 185(3) MeV. 

These allow for significant improvements in predictions for 
SM rates. For example, updating~\cite{Buras:2012ru} with 
the world-average for $f_{B_s}$ above and our result for 
$f_{B^0}$ (Eq.~\ref{eq:fmoreres}) we obtain: 
\begin{eqnarray}
\mathrm{Br}(B_s \rightarrow \mu^+\mu^-) &=& 3.17\pm 0.15 \pm 0.09 \times 10^{-9} \nonumber \\
\mathrm{Br}(B_d \rightarrow \mu^+\mu^-) &=& 1.05 \pm 0.05 \pm 0.05 \times 10^{-10} 
\label{eq:br}
\end{eqnarray}
where the second error from $f_{B_q}$ has been halved and is no longer
larger than other sources of error such as $V_{tb}^*V_{tq}$. Note that this 
is the flavor-averaged branching fraction at $t=0$; the time-integrated 
result would be increased by 10\% in the $B_s$ case (to $3.47(19) \times 10^{-9}$)
to allow for the width difference 
of the two eigenstates~\cite{DeBruyn:2012wk, lhcb-delta}. The current 
experimental results~\cite{lhcb:2012ct} for $B_s \rightarrow \mu^+\mu^-$ agree with this prediction. 

From the world-average $f_{B^+}$ above we also obtain the Standard Model rate:
\begin{equation}
\frac{1}{|V_{ub}|^2}\mathrm{Br}(B^+ \rightarrow \tau \nu) = 6.05(20),
\end{equation}
with 3\% accuracy.  Calculations of matrix elements
 for $B_s/B$ mixing with physical $u/d$ quarks are now underway.

\noindent{{\bf Acknowledgements}} We are grateful to the MILC collaboration for the use of their 
gauge configurations and to B. Chakraborty, J. Koponen and P. Lepage 
for useful discussions.  
The results described here were obtained using the Darwin Supercomputer 
of the University of Cambridge High Performance 
Computing Service as part of STFC's DiRAC facility. 
This work was funded by STFC and the US DOE. 


\bibliography{hl_bib}

\end{document}